\documentclass[%
aps,nofootinbib,notitlepage,superscriptaddress,10Limberkipt,prd, twocolumn,
 amsmath,amssymb
]{revtex4-2}

\usepackage[caption=false]{subfig}
\usepackage{braket}
\usepackage{float}
\usepackage{lipsum}
\usepackage{graphicx}
\usepackage{dcolumn}
\usepackage{bm}
\usepackage{natbib}
\usepackage{hyperref}
\usepackage{tcolorbox}
\hypersetup{
	colorlinks = true,
    linkcolor = Maroon,
    urlcolor  = Maroon,
    citecolor = Maroon
}
\usepackage{graphicx}
\usepackage{microtype}
\usepackage{verbatim}
\usepackage[amssymb]{SIunits}
\usepackage{tabularx}
\usepackage[dvipsnames]{xcolor}
\usepackage{tikz}

\newcommand{\columbia}{Department of Physics, Columbia University, New York, NY, USA 10027}

\begin{document}

\preprint{APS/123-QED}

\title{A 2\% determination of $N_{\rm eff}$ from primordial element abundance, cosmic microwave background, and baryon acoustic oscillation measurements}
\author{Samuel~Goldstein}
\email{sjg2215@columbia.edu}
\affiliation{\columbia}

\author{J.~Colin Hill}
\email{jch2200@columbia.edu}
\affiliation{\columbia}

\begin{abstract}
  \noindent We present a new constraint on the effective number of relativistic species in the early universe, $N_{\rm eff}$, by combining recent primordial helium abundance measurements from the Large Binocular Telescope $Y_p$ Project with primordial deuterium abundance data, cosmic microwave background (CMB) observations from \emph{Planck}, the Atacama Cosmology Telescope, and the South Pole Telescope, and baryon acoustic oscillation (BAO) data from the Dark Energy Spectroscopic Instrument, yielding $N_{\rm eff}=2.990\pm0.070$ (68\% C.L.). This is the tightest constraint on $N_{\rm eff}$ to date, and is in excellent agreement with the standard model prediction of $N_{\rm eff}=3.044$. Furthermore, we constrain excess contributions to $N_{\rm eff}$ beyond the three neutrino species, finding $\Delta N_{\rm eff}<0.107$ (95\% C.L.).  This bound nearly approaches the minimum contribution to $\Delta N_{\rm eff}$ from a light spin-3/2 particle that decoupled at any time after inflation ended.  Our baseline analysis does not include large-scale \emph{Planck} polarization information, enabling a fully consistent combination of state-of-the-art CMB and BAO measurements. As a byproduct, we show that current $N_{\rm eff}$ bounds are essentially insensitive to the inclusion or exclusion of optical depth constraints inferred from large-scale CMB polarization data, making $N_{\rm eff}$ highly robust in this regard.  Our constraints place stringent limits on light particles in the early Universe and on a broad range of models aimed at increasing the CMB-inferred value of the Hubble constant.
\end{abstract}

\maketitle

\section{Introduction}\label{sec:introduction}

\noindent The effective number of relativistic species, $N_{\rm eff}$, quantifies contributions from particles other than photons to the radiation energy density in the early Universe. It is defined by (see, \emph{e.g.}, Ref.~\cite{Baumann:2022mni})
\begin{equation}
    \rho_{\rm rad}=\left[1+\frac{7}{8}\left(\frac{4}{11}\right)^{4/3}N_{\rm eff}\right]\rho_{\gamma},
\end{equation}
where $\rho_{\rm rad}$ and $\rho_{\gamma}$ define the total radiation energy density and photon energy density, respectively. In the Standard Model, the three neutrino species contribute $N_{\rm eff}=3.044$, where the small excess over 3 arises primarily from non-instantaneous neutrino decoupling and the near-contemporaneous annihilation of electrons and positrons~\cite{Mangano:2001iu, Mangano:2005cc, deSalas:2016ztq, Akita:2020szl, Froustey:2020mcq, Bennett:2020zkv, Drewes:2024wbw}. However, many beyond-Standard-Model (BSM) scenarios predict new light degrees of freedom that would contribute to $N_{\rm eff}$, including axions, dark photons, sterile neutrinos, and high-frequency gravitational waves~\cite{Brust:2013ova,Salvio:2013iaa,Kawasaki:2015ofa,Baumann:2016wac,Abazajian:2001nj,Strumia:2006db,Boyarsky:2009ix,Boyle:2007zx,Stewart:2007fu,Ackerman:2008kmp,Kaplan:2011yj,Cyr-Racine:2012tfp,Cadamuro:2010cz,Weinberg:2013kea,Nakayama:2010vs,Arkani-Hamed:2016rle,CMB-S4:2016ple}; alternatively, finding $N_{\rm eff} < 3.044$ may indicate photon heating after neutrino decoupling~\cite{Steigman:2013yua,Boehm:2013jpa} or a low post-inflationary reheating temperature~\cite{Hasegawa:2019jsa,Abazajian:2023reo,Barbieri:2025moq}. In general, $N_{\rm eff}$ is a powerful, wide-ranging probe of new physics in the early Universe.

The presence of additional radiation directly affects the expansion history and perturbation evolution in the early Universe, hence $N_{\rm eff}$ modifies the cosmic microwave background (CMB)~\cite{Bashinsky:2003tk, Baumann:2015rya} and alters the abundances of light elements forged during Big Bang nucleosynthesis (BBN)~\cite{Steigman:1977kc}. Consequently, primordial element abundance measurements and CMB observations can be used to place stringent constraints on $N_{\rm eff}$.  Recently, by combining a new measurement of the primordial helium abundance with \emph{Planck} CMB observations, the Large Binocular Telescope (LBT) $Y_P$ project reported the tightest constraint on $N_{\rm eff}$ in the literature: $N_{\rm eff}=2.968\pm0.083$~\cite{LBT_PartI, LBT_PartII, LBT_PartIII, LBT_PartIV, LBT_PartV}.\footnote{The LBT result is reported in terms of the effective number of neutrino species, $N_{\nu}=2.925\pm0.082$, which we rescale to $N_{\rm eff}$ by multiplying by $3.044/3.$ Unless otherwise stated, all constraints are reported in terms of the posterior mean and the 68\% two-tailed C.L.}  Meanwhile, new high-resolution CMB temperature and polarization observations from the Atacama Cosmology Telescope (ACT)~\cite{AtacamaCosmologyTelescope:2025vnj,ACT:2025fju} and the South Pole Telescope (SPT)~\cite{SPT-3G:2025bzu} have been used to derive constraints on $N_{\rm eff}$, improving the bound from \emph{Planck} alone by nearly a factor of two~\cite{ACT:2025tim, SPT-3G:2025bzu}.  Including baryon acoustic oscillation (BAO) data from the Dark Energy Spectroscopic Instrument (DESI) slightly tightens the constraint further due to breaking of parameter degeneracies, yielding $N_{\rm eff}=2.86\pm0.13$ from \emph{Planck} + ACT + DESI DR1~\cite{ACT:2025tim}.

In this brief paper, we assess the consistency of the $N_{\rm eff}$ constraints from these various datasets and, ultimately, combine them to derive the most precise constraint on $N_{\rm eff}$ to date, $N_{\rm eff}=2.990\pm 0.070$, improving upon the LBT + \emph{Planck} constraint by $\approx 15\%$.

\section{Datasets and methodology}\label{sec:datasets_method}
\noindent In this section, we describe the datasets and computational codes used in this work. We use the recent measurement of the primordial helium nucleon fraction by the LBT $Y_P$ project~\cite{LBT_PartV},\footnote{Throughout, we follow the \texttt{camb} convention and define $Y_{P}$ ($Y_{\rm He}$) as the primordial helium nucleon (mass) fraction. These differ by $\sim0.5\%$, due to the binding energy of helium. We note that astrophysical measurements of the primordial helium ``mass fraction" typically correspond to the nucleon fraction (see, \emph{e.g.}, Appendix A of \cite{Fields:2019pfx}).}
\begin{equation} \label{eq.LBTYP}
    Y_{P}^{\rm LBT26}=0.2458\pm0.0013 \,.
\end{equation}
The LBT measurement is over a factor of two more constraining than the 2024 Particle Data Group (PDG) value~\cite{ParticleDataGroup:2024cfk} that was employed as an external constraint in the recent ACT Data Release 6 (DR6) CMB power spectrum analysis~\cite{ACT:2025tim}. We implement Eq.~\eqref{eq.LBTYP} as a Gaussian likelihood on $Y_P$ in our analysis. Additionally, following Ref.~\cite{ACT:2025tim}, we include measurements of the primordial deuterium abundance as an external Gaussian likelihood with 
\begin{equation}
    {\rm D/H}\vert_P\times 10^{5}=2.547\pm 0.104,
\end{equation} 
which is based on the consensus value in the 2024 PDG report, accounting for both observational and theoretical uncertainties~\citep{Cooke:2013cba,Cooke:2016rky,Riemer-Sorensen:2014aoa,Balashev:2015hoe,Riemer-Sorensen:2017pey}. Given the current uncertainties, the deuterium abundance has a negligible impact on constraints on $N_{\rm eff}$ when CMB data are included; however, both the primordial helium and deuterium abundances are essential for constraining $N_{\rm eff}$ using BBN alone, due to the differing sensitivities of the element abundances to $N_{\rm eff}$ and the baryon density $\Omega_b h^2$. We refer to the helium and deuterium abundance datasets as $Y_{P}^{\rm LBT26}$ and ${\rm D}$, respectively.

We include measurements of the CMB temperature and polarization power spectra, as well as the CMB lensing potential power spectrum. For reference, we use the \emph{Planck} 2018 high-$\ell$ (TT+TE+EE) and low-$\ell$ (TT+EE) likelihoods together with the \emph{Planck} CMB 2018 lensing likelihood~\cite{Aghanim:2019ame, Aghanim:2018oex}. In our baseline analysis, we conservatively exclude \emph{Planck} low-$\ell$ EE measurements because of the mild inconsistency between the CMB and BAO inferences of $\Omega_m$ and $H_0r_d$ in $\Lambda$CDM (and $N_{\rm eff}$) cosmologies. As shown in Refs.~\cite{Loverde:2024nfi,  Allali:2025wwi, Sailer:2025lxj, Jhaveri:2025neg}, this inconsistency can be significantly alleviated by relaxing the constraint on the optical depth to reionization, $\tau$, which is set by the \emph{Planck} large-scale polarization data. Thus, by excluding low-$\ell$ EE measurements, we can robustly combine CMB and BAO observations. For comparison, we include results using the \texttt{Sroll2} low-$\ell$ EE likelihood~\cite{Pagano:2019tci}. As we shall see, the low-$\ell$ EE data have a negligible impact on our final $N_{\rm eff}$ constraint. We discuss the impact of low-$\ell$ polarization information on $N_{\rm eff}$ constraints in more detail in Appendix~\ref{app:lowl_EE}.

We also include high-resolution temperature and polarization data from ACT DR6~\cite{AtacamaCosmologyTelescope:2025vnj,ACT:2025fju,ACT:2025tim,Beringue:2025bur} and the SPT-3G D1~\cite{SPT-3G:2025bzu}. Following the ACT DR6 analysis, we combine the \emph{Planck} and ACT primary CMB likelihoods using the ``P-ACT" combination, which restricts the \emph{Planck} primary spectra to $\ell < 1000$ for TT and $\ell < 600$ for TE and EE. For CMB lensing, with the exception of our \emph{Planck}-only analysis, which we include to reproduce the results from Refs.~\cite{Aghanim:2018eyx, LBT_PartV}, we use the combined \emph{Planck}, ACT, and SPT CMB lensing likelihood developed in Ref.~\cite{ACT:2025qjh}. We denote our combined \emph{Planck}, ACT, and SPT CMB likelihood as CMB-PAS.
\footnote{Our CMB-PAS dataset combination is similar to the CMB-SPA combination used in Ref.~\cite{SPT-3G:2025bzu}, with the key difference that, when including large-scale polarization measurements, we use the \texttt{Sroll2} low-$\ell$ EE likelihood, whereas the CMB-SPA combination imposes a Gaussian prior on the optical depth to reionization, $\tau\sim \mathcal{N}(0.051, 0.006)$, based on Ref.~\cite{Planck:2020olo}.}

Finally, we include the latest BAO distance measurements from the second data release (DR2) of the DESI survey~\cite{DESI:2025zpo,DESI:2025zgx}. Specifically, we use the combined DESI DR2 BAO sample, which includes transverse and line-of-sight BAO measurements from galaxies, quasars, and the Lyman-$\alpha$ forest, spanning redshifts $0.295\leq z\leq 2.33$.

We compute theoretical predictions using the~\texttt{camb}~\cite{Lewis:1999bs} Einstein-Boltzmann code, the \texttt{CosmoRec} recombination code~\cite{COSMOREC}, and the \texttt{PRyMordial} BBN code~\cite{Burns:2023sgx}. We fix the total neutrino mass to $\sum m_\nu=0.06$ eV, assuming a single massive eigenstate. We list our \texttt{camb} settings in Appendix~\ref{app:camb_settings}. We use the \texttt{candl} implementation of the SPT-3G D1 likelihood~\cite{Balkenhol:2024sbv}.\footnote{\href{https://github.com/SouthPoleTelescope/spt_candl_data}{https://github.com/SouthPoleTelescope/spt\_candl\_data}} We sample from the parameter posterior distributions using \texttt{Cobaya}~\cite{Torrado:2020dgo},\footnote{\href{https://cobaya.readthedocs.io/en/latest/}{https://cobaya.readthedocs.io/en/latest/}} adopting the following uniform priors: $0.017\leq \Omega_bh^2\leq 0.027$, $0.09\leq \Omega_c h^2\leq 0.15,$ $0.01038\leq \theta_{\rm MC}\leq 0.01044$, $2.6 \leq \ln(10^{10}A_s)\leq 3.5$, $0.85\leq n_s\leq 1.1,$ and $0.0\leq \tau\leq 0.15$. For our fiducial constraints on $N_{\rm eff}$, we impose a uniform prior of $0.1\leq N_{\rm eff}\leq 6.0$. When placing constraints on the \emph{excess} number of relativistic degrees of freedom $\Delta N_{\rm eff}\equiv N_{\rm eff}-3.044$, we impose a prior of $3.044 \leq N_{\rm eff}\leq 6.0$. In addition to the cosmological parameters, we vary all recommended nuisance parameters for each dataset using the priors from Refs.~\cite{ACT:2025fju, SPT-3G:2025bzu}. We analyze chains using the \texttt{getdist} package~\cite{Lewis:2019xzd}, and assess convergence using the Gelman-Rubin statistic~\cite{gelman1992}, requiring $|R-1|< 0.02$.

\begin{figure}[!t]
\centering
\includegraphics[width=0.99\linewidth]{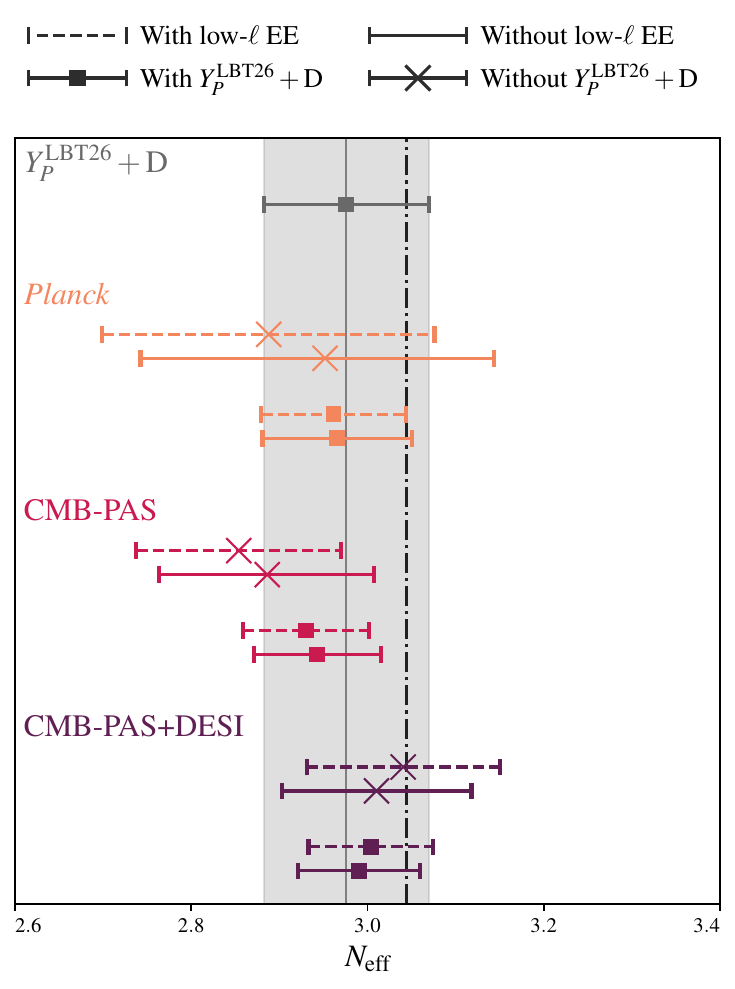}
\caption{Comparison of $N_{\rm eff}$ constraints from dataset combinations considered here. For each dataset, we show the one-dimensional marginalized posterior mean and the 68\% two-tailed confidence limits. The gray constraint uses only primordial helium and deuterium abundance measurements. For the remaining datasets, the $\times$ and the square denote constraints excluding and including primordial helium and deuterium abundance measurements, respectively. The measurements with solid (dashed) error bars exclude (include) \emph{Planck} low-$\ell$ EE measurements. The vertical dot-dashed line shows the Standard Model prediction, $N_{\rm eff}=3.044.$}
\label{fig:Neff_1D_posterior}
\end{figure}

\section{Results}\label{sec:results}

\noindent In Fig.~\ref{fig:Neff_1D_posterior}, we show the one-dimensional marginalized posteriors for $N_{\rm eff}$ obtained from various dataset combinations considered in this work. The markers denote the posterior mean and the error bars indicate the 68\% two-tailed C.L. The solid (dashed) error bars exclude (include) large-scale \emph{Planck} polarization data. Using only primordial helium and deuterium abundance measurements, we find $N_{\rm eff}=2.976\pm 0.093$ (gray region), fully consistent with the ``BBN+$Y_p$+D" constraint in Ref.~\cite{LBT_PartV}. Using only \emph{Planck} CMB data (including low-$\ell$ EE), we find $N_{\rm eff}=2.89\pm0.19$ (orange $\times$), matching Ref.~\cite{Aghanim:2018eyx}. Combining the LBT measurement and deuterium abundance priors with \emph{Planck} 2018 CMB observations, we find $N_{\rm eff}=2.961\pm 0.082$, again consistent with Ref.~\cite{LBT_PartV}.

Having reproduced the tightest $N_{\rm eff}$ constraint currently in the literature, we now turn to the main purpose of this brief paper: combining the new helium abundance measurements with the latest CMB and BAO observations. Using primary CMB and lensing measurements from \emph{Planck}, ACT, and SPT, we find $N_{\rm eff}=2.85\pm0.12$ and $N_{\rm eff}=2.89\pm0.12$ with and without low-$\ell$ EE measurements, respectively (dark pink $\times$), consistent with Refs.~\cite{ACT:2025tim, SPT-3G:2025bzu}.\footnote{Our CMB-PAS constraint including low-$\ell$ EE data is slightly higher than the CMB-SPA result, $N_{\rm eff}=2.81\pm0.12$, in Ref.~\cite{SPT-3G:2025bzu}, as we use the \texttt{Sroll2} likelihood rather than a prior of $\tau=0.051\pm0.006.$} Thus, low-$\ell$ EE data have little impact on current CMB-derived $N_{\rm eff}$ bounds.

The CMB-only constraints are entirely consistent with the primordial abundance constraint; there is no evidence for a change in the number of relativistic species between BBN at redshift $z \approx 10^8$-$10^9$ and CMB decoupling at $z \approx 1100$. Moreover, the CMB constraints are competitive with that derived from the light element abundance data. Therefore, we combine these datasets to obtain an even tighter joint constraint, yielding $N_{\rm eff}=2.930\pm0.072$ and $2.943\pm 0.072$ with and without low-$\ell$ EE data, respectively (dark pink squares).

Finally, we combine the CMB-PAS likelihoods with DESI DR2 BAO observations. As described in Appendix~\ref{app:lowl_EE} of this work, if low-$\ell$ EE data are included, the CMB and BAO data are inconsistent at $\sim 3\sigma$ in the $(H_0r_d, \Omega_m, N_{\rm eff})$ parameter space. However, without low-$\ell$ EE data, the CMB and BAO data are fully consistent, and, hence, can be combined. Combining CMB-PAS (without low-$\ell$ EE) with DESI DR2 BAO observations, we find $N_{\rm eff}=3.01 \pm 0.11$ (dark purple $\times$), again consistent with the abundance-inferred $N_{\rm eff}$ constraint from BBN. We thus combine all of these datasets (except low-$\ell$ EE), yielding\footnote{For completeness, we also analyze DESI DR2 BAO with CMB-PAS including low-$\ell$ EE measurements, yielding $N_{\rm eff}=3.004\pm0.071$ and $N_\mathrm{eff} = 3.04\pm 0.11$ with and without abundance information, respectively. While these constraints are completely consistent with our results that exclude the low-$\ell$ EE data, we note that there is a mild inconsistency between CMB and BAO data when the low-$\ell$ EE measurements are included.  Nevertheless, the inclusion or exclusion of low-$\ell$ EE data has essentially no impact on the $N_{\rm eff}$ constraint, even from CMB-PAS data alone.}
\begin{equation} \label{eq.Neffconstraint}
    N_{\rm eff}=2.990\pm 0.070 \,.
\end{equation}
This is the main result of this brief paper: a 2\% determination of $N_{\rm eff}$ that is in excellent agreement with the Standard Model prediction, $N_{\rm eff}=3.044$. This is the tightest bound on $N_{\rm eff}$ obtained to date, improving by $\approx 15\%$ over the LBT+\emph{Planck} bound~\cite{LBT_PartV}. 

We now make several comments regarding this result. First, in the absence of primordial abundance information, including DESI data shifts the central value of $N_{\rm eff}$ upward by $\Delta N_{\rm eff} \sim 0.1$ (when abundances are included, the shift is reduced to $\sim 0.05$). Quantitatively, this is a slightly surprising shift given that the error bar is not reduced significantly by DESI (see, e.g.,~\cite{Gratton:2019fru} for metrics to quantify such parameter shifts). Ultimately, this shift in $N_{\rm eff}$ is related to the mild inconsistency between the matter fraction $\Omega_m$ values inferred by DESI and the primary CMB, as discussed in, \emph{e.g.}, Ref.~\cite{ACT:2025tim} and Appendix~\ref{app:lowl_EE} of this work. 

Second, our results illustrate that low-$\ell$ polarization data have negligible impact on current CMB-derived constraints on $N_{\rm eff}$ (see also, \emph{e.g.}, Ref.~\cite{Giare:2023ejv} for an earlier analysis). Thus, we can construct a near-optimal and completely consistent combination of CMB and BAO data for $N_{\rm eff}$ constraints by simply removing low-$\ell$ EE data. Looking ahead, this result suggests that, unlike constraints on the neutrino mass, constraints on $N_{\rm eff}$ from high-resolution CMB observations will not be limited by uncertainties in $\tau$, which is predominately constrained by low-$\ell$ CMB polarization measurements.

\begin{figure}[!t]
\centering
\includegraphics[width=0.99\linewidth]{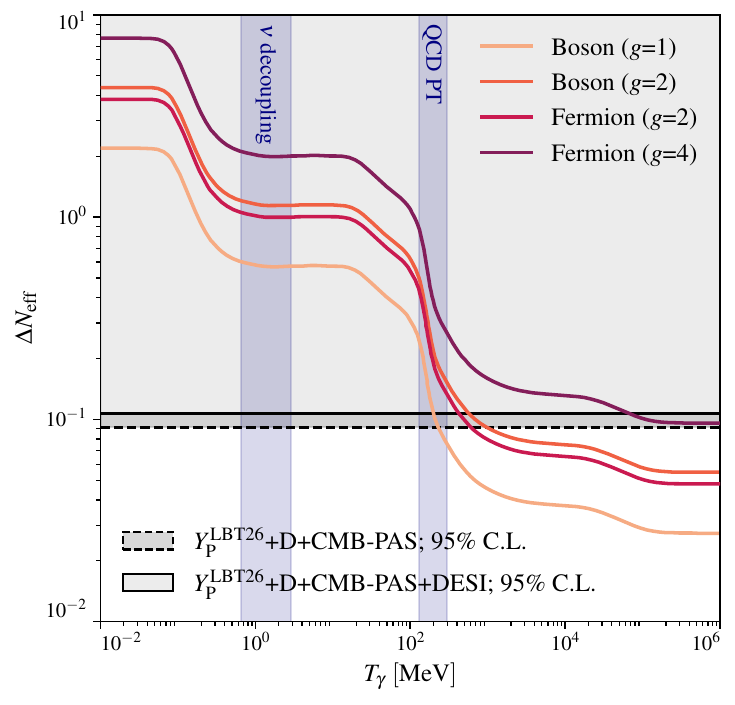}
\caption{Constraints on additional relativistic degrees of freedom in the early Universe. The light (dark) gray shaded region denotes $\Delta N_{\rm eff} > 0.107$ ($0.0913$), which is excluded at the 95\% C.L. using primordial abundance, CMB, and BAO data (primordial abundance and CMB data). In both cases, we omit low-$\ell$ EE data. The solid curves show theoretical predictions for $\Delta N_{\rm eff}$ as a function of the particle decoupling temperature for bosons and fermions with different numbers of internal degrees of freedom, $g$. The vertical blue bands denote the approximate regions of neutrino decoupling and the QCD phase transition. This figure is based on Refs.~\cite{Aghanim:2018eyx, SimonsObservatory:2018koc, ACT:2025tim}.}
\label{fig:delta_Neff_exclusion}
\end{figure}

We now turn to the physical implications of our results. First, we can convert our excess $\Delta N_{\rm eff}$ bounds into constraints on any additional relativistic species in the early Universe that were in thermal equilibrium with the Standard Model. Combining the primordial abundance measurements with CMB (without low-$\ell$ EE) and BAO observations, and imposing a prior of $\Delta N_{\rm eff}>0$, we find
\begin{equation}
    \Delta N_{\rm eff}<0.107~(95\%~{\rm C.L.}),
\end{equation}
which is the tightest constraint on $\Delta N_{\rm eff}$ to date. If we exclude the DESI BAO observations, the constraint slightly tightens to $\Delta N_{\rm eff} < 0.0913$.\footnote{Including low-$\ell$ EE data, we find $\Delta N_{\rm eff}<0.116$ and $\Delta N_{\rm eff}<0.0876$ with and without DESI DR2 BAO data, respectively. } 

In Fig.~\ref{fig:delta_Neff_exclusion}, we compare our 95\% exclusion regions with theoretical predictions for $\Delta N_{\rm eff}$ as a function of the particle decoupling temperature, $T_{\gamma}$, and the number of internal degrees of freedom, $g$, from Ref.~\cite{Borsanyi:2016ksw}. For either dataset combination shown, we exclude any type of new light particle that decoupled after the QCD phase transition at over 5$\sigma$. The constraints are even tighter at the later epoch of neutrino decoupling. Furthermore, since the number of internal degrees of freedom is directly related to the spin of a particle ($g=2s+1$), the limit without DESI data excludes any light spin-3/2 particle that was ever in thermal equilibrium with the primordial plasma all the way back to reheating at just over $2\sigma$. 

Second, we emphasize that the consistency between the $N_{\rm eff}$ values inferred from BBN and from the CMB (and BAO) places strong constraints on BSM scenarios that affect cosmic evolution between BBN and recombination, such as the presence of a BSM particle that becomes non-relativistic after BBN but before recombination (see Ref.~\cite{Yeh:2022heq} for a detailed discussion of scenarios where $N_{\rm eff}^{\rm BBN}\neq N_{\rm eff}^{\rm CMB}$). In the future, it would be interesting to perform a joint analysis of $N_{\rm eff}^{\rm BBN}$ and $N_{\rm eff}^{\rm CMB}$ in the context of such models using the datasets considered here.

Finally, our constraints have important implications for the ongoing debate regarding the value of the Hubble constant. Combining primordial abundances, CMB-PAS (without low-$\ell$ EE), and DESI DR2 BAO measurements, we find $H_0=68.16\pm0.50$ km/s/Mpc in a $\Lambda$CDM+$N_{\rm eff}$ cosmology, far below the latest measurement from the SH0ES collaboration, $H_0=73.17\pm0.86$~km/s/Mpc~\cite{Riess:2021jrx, Breuval:2024lsv} or the recent result from the Local Distance Network, $H_0 = 73.50 \pm 0.81$~km/s/Mpc~\cite{H0DN:2025lyy}.\footnote{We note that the CMB and BAO-inferred $H_0$ constraints presented in this work are statistically consistent with the CCHP measurement, $H_0=70.39\pm1.94$ km/s/Mpc~\cite{Freedman:2024eph}.} Formally, our $H_0$ constraint for the $\Lambda$CDM+$N_{\rm eff}$ scenario is in $5\sigma$ tension with the SH0ES measurement, firmly excluding free-streaming dark radiation as a solution to the Hubble constant discrepancy. Moreover, the strength of this exclusion suggests that the datasets considered here likely rule out many related extensions to $\Lambda$CDM that aim to increase $H_0$, such as strongly interacting dark radiation and variants thereof~\cite{Jeong_Takahashi_2013,Buen-Abad_2015,Cyr-Racine_2016,Lesgourgues_2016,Aloni:2021eaq,Joseph_2023,Buen-Abad_2023,Rubira_2023,Schoneberg_2023,Zhou_Weiner_2024}.  Models attempting to (partially) evade our stringent bound likely must populate the dark radiation sector after the BBN epoch in order to avoid the very strong LBT-derived $N_{\rm eff}^{\rm BBN}$ constraint (see, \textit{e.g.},~\cite{Aloni:2021eaq,Aloni:2023tff,Aloni:2024eay,Garny:2025kqj} for models of this sort).  However, the CMB-PAS + BAO constraint on $N_{\rm eff}^{\rm CMB}$ is also very tight, thus leaving very little wiggle room for the production of dark radiation between the BBN and CMB epochs.  Indeed, a dedicated study of many such BSM models using ACT DR6, \emph{Planck}, and DESI DR1 BAO data was performed in Ref.~\cite{ACT:2025tim}, yielding very stringent constraints and no hint of a higher $H_0$ value.  We leave a detailed study of such models using joint BBN, CMB-PAS, and DESI DR2 BAO data to future work.

\section{Conclusions}\label{sec:results}
\noindent In this brief paper, we have presented a new constraint on the effective number of relativistic species, $N_{\rm eff}$, by combining state-of-the-art measurements of the primordial helium and deuterium abundances, the cosmic microwave background, and baryon acoustic oscillations, yielding $N_{\rm eff}=2.990\pm0.070$ (Eq.~\eqref{eq.Neffconstraint}).

Our 2\% determination of $N_{\rm eff}$ is the most precise constraint to date, and in excellent agreement with the Standard Model prediction. Moreover, we constrained excess contributions to the effective number of relativistic species, $\Delta N_{\rm eff}<0.107$ (95\% C.L.), excluding any new light particle that decoupled from the thermal bath after the QCD phase transition at over 5$\sigma$. We also found that the Hubble constant discrepancy persists at over 5$\sigma$ for $N_{\rm eff}$ cosmologies.

Our stringent $N_{\rm eff}$ constraint demonstrates the power of combining modern astrophysical and cosmological measurements. To reliably perform such a combination, we assessed the consistency of $N_{\rm eff}$ constraints from primordial abundances, the CMB, and BAO, finding them to be in excellent agreement, with the exception of a mild tension between the CMB and BAO when low-$\ell$ EE data are included, which we thus excluded from our baseline analysis with a negligible loss in constraining power on $N_{\rm eff}$. We note that although we have held the neutrino mass sum fixed to 0.06~eV in this analysis, our $N_{\rm eff}$ constraint should be highly insensitive to $\sum m_\nu$ because it is dominated by the new LBT $Y_P$ measurement, which is insensitive to $\sum m_\nu$, and because current CMB-derived constraints on $N_{\rm eff}$ are largely independent of $\sum m_\nu$~\cite{ACT:2025tim}. In the near future, improved astrophysical determinations of the primordial helium fraction, high-resolution CMB observations from the Simons Observatory~\cite{SimonsObservatory:2018koc,SimonsObservatory:2025wwn}, and next-generation large-scale structure surveys such as DESI-II~\cite{DESI:2022lza} and \emph{Euclid}~\cite{Euclid:2024yrr} will increase our sensitivity to $N_{\rm eff}$, potentially providing evidence for new light particles in the early Universe.


\acknowledgments 

\noindent We thank Brian Fields for helpful conversations. We thank Bruce Partridge and Ed Wollack for useful feedback on this manuscript. SG and JCH acknowledge support from DOE grant HEP DE-SC0011941 and NSF grant AST-2307727.  JCH also acknowledges support from NASA grant 80NSSC23K0463 [ADAP] and the Sloan Foundation.  The authors acknowledge the Texas Advanced Computing Center (TACC)\footnote{\href{http://www.tacc.utexas.edu}{http://www.tacc.utexas.edu}} at The University of Texas at Austin for providing computational resources that have contributed to the research results reported within this paper. We acknowledge computing resources from Columbia University's Shared Research Computing Facility project, which is supported by NIH Research Facility Improvement Grant 1G20RR030893-01, and associated funds from the New York State Empire State Development, Division of Science Technology and Innovation (NYSTAR) Contract C090171, both awarded April 15, 2010. 
\noindent 
 \bibliographystyle{apsrev4-1}
\bibliography{biblio.bib}

\clearpage
\pagebreak
\phantomsection

\onecolumngrid

\appendix

\begin{figure*}
    \centering
    \includegraphics[width=0.95\linewidth]{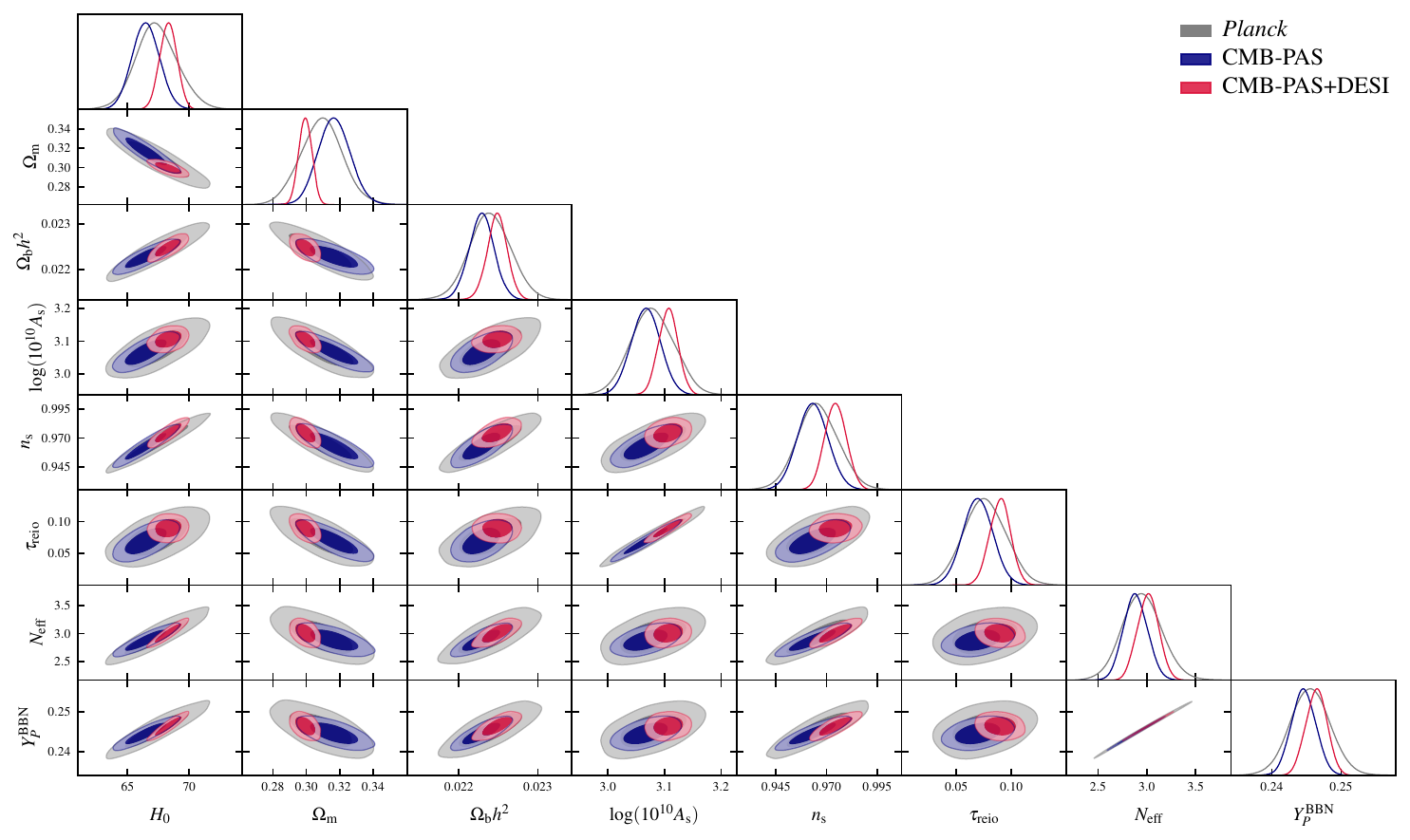}
    \caption{Two-dimensional marginalized posterior for the six $\Lambda$CDM parameters, as well as $N_{\rm eff}$ and $Y_P^{\rm BBN}$, for the main datasets considered here. Note that these results do not include low-$\ell$ EE data. This figure does not impose any external priors on primordial elemental abundances (see Fig.~\ref{fig:full_posterior_with_abundance_info}).}
    \label{fig:full_posterior_no_abundance_info}
\end{figure*}

\begin{figure*}
    \centering
    \includegraphics[width=0.95\linewidth]{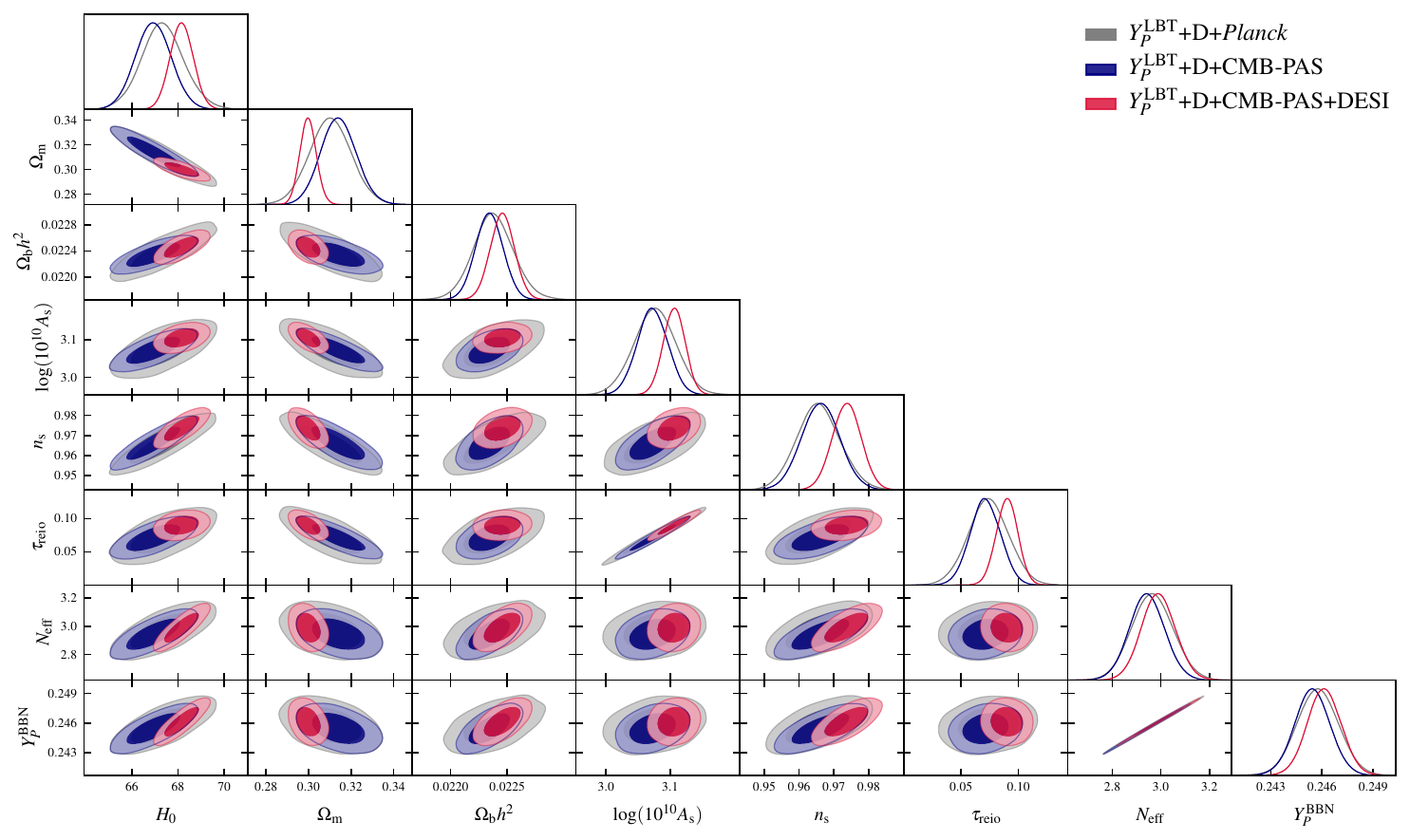}
    \caption{Same as Fig.~\ref{fig:full_posterior_no_abundance_info}, except including measurements of the primordial helium and deuterium abundance.}
    \label{fig:full_posterior_with_abundance_info}
\end{figure*}

\clearpage
\pagebreak
\section{Two-dimensional marginalized posterior distributions and table of parameter constraints}\label{app:param_constraints_and_tables}

\noindent Figures~\ref{fig:full_posterior_no_abundance_info} and~\ref{fig:full_posterior_with_abundance_info} show the two-dimensional marginalized posterior distributions of the six $\Lambda$CDM parameters, $N_{\rm eff}$, and $Y_P^{\rm BBN}$ for the dataset combinations considered in this work, excluding and including primordial helium and deuterium abundance measurements, respectively. Following our baseline analysis choices, these results do not include low-$\ell$ EE data. Table~\ref{table:baseline_param_constraints} shows the marginalized posterior mean and 68\% two-tailed confidence limit for these dataset combinations.

\begin{table}[!h]
\centering
\resizebox{\textwidth}{!}{
\begin{tabular}{|l|c|c|c|c|c|c|c|c|}
\multicolumn{9}{c}{Without abundance information} \\
\hline
Dataset & $H_0$ & $\Omega_m$ & $\Omega_b h^2$ & $\tau$ & $\log A_s$ & $n_s$ & $N_{\rm eff}$ & $Y_P$ \\
\hline
    \emph{Planck} & $67.4^{+1.6}_{-1.8}$ & $0.309 \pm 0.013$ & $0.02240 \pm 0.00026$ & $0.076 \pm 0.020$ & $3.078 \pm 0.040$ & $0.966^{+0.010}_{-0.010}$ & $2.95^{+0.19}_{-0.21}$ & $0.2455 \pm 0.0028$ \\
    CMB+PAS & $66.5 \pm 1.1$ & $0.3166 \pm 0.0096$ & $0.02229 \pm 0.00016$ & $0.069 \pm 0.014$ & $3.067 \pm 0.026$ & $0.9633 \pm 0.0075$ & $2.89 \pm 0.12$ & $0.2446 \pm 0.0018$ \\
    CMB+PAS+DESI & $68.30 \pm 0.71$ & $0.2993 \pm 0.0040$ & $0.02248 \pm 0.00013$ & $0.0904 \pm 0.0092$ & $3.107 \pm 0.017$ & $0.9745 \pm 0.0053$ & $3.01 \pm 0.11$ & $0.2464 \pm 0.0015$ \\
\hline
\end{tabular}
}

\vspace{0.5em}

\resizebox{\textwidth}{!}{
\begin{tabular}{|l|c|c|c|c|c|c|c|c|}
\multicolumn{9}{c}{With abundance information} \\
\hline
Dataset & $H_0$ & $\Omega_m$ & $\Omega_b h^2$ & $\tau$ & $\log A_s$ & $n_s$ & $N_{\rm eff}$ & $Y_P$ \\
\hline
    \emph{Planck} & $67.34 \pm 0.91$ & $0.3103 \pm 0.0097$ & $0.02238 \pm 0.00018$ & $0.074 \pm 0.018$ & $3.076 \pm 0.032$ & $0.9656 \pm 0.0062$ & $2.965 \pm 0.085$ & $0.2458 \pm 0.0012$ \\
    CMB+PAS & $66.93 \pm 0.77$ & $0.3141 \pm 0.0083$ & $0.02234 \pm 0.00012$ & $0.071 \pm 0.013$ & $3.072 \pm 0.024$ & $0.9662 \pm 0.0054$ & $2.943 \pm 0.072$ & $0.2454 \pm 0.0010$ \\
    CMB+PAS+DESI & $68.16 \pm 0.50$ & $0.2998 \pm 0.0038$ & $0.02246 \pm 0.00011$ & $0.0901 \pm 0.0094$ & $3.106 \pm 0.017$ & $0.9737 \pm 0.0041$ & $2.990 \pm 0.070$ & $0.24613 \pm 0.00098$ \\
\hline
\end{tabular}
}
\caption{Constraints on the key cosmological parameters for various dataset combinations considered in this work. For each parameter, we show the posterior mean and 68\% two-tailed confidence limit. The top table shows constraints derived only using CMB and BAO observations. The bottom table includes astrophysical measurements of the primordial helium and deuterium abundance. These results do not include \emph{Planck} low-$\ell$ EE measurements.}
\label{table:baseline_param_constraints}
\end{table}

\clearpage\pagebreak

\section{Impact of low-$\ell$ polarization measurements on $N_{\rm eff}$ constraints}\label{app:lowl_EE}

\begin{figure*}[!t]
    \centering
    \includegraphics[width=0.95\linewidth]{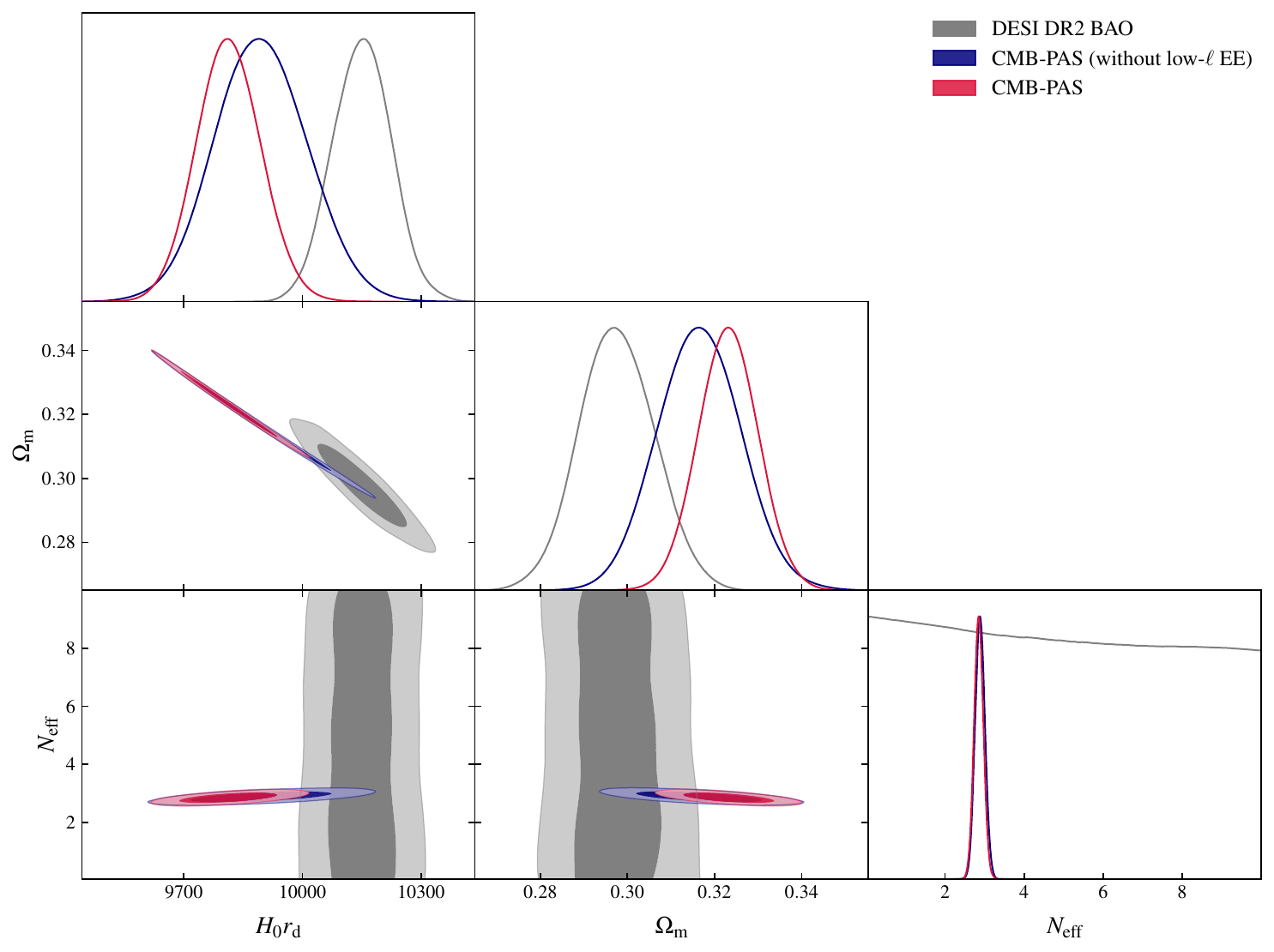}
        \caption{Comparison of constraints on $H_0r_d$, $\Omega_m$, and $N_{\rm eff}$ inferred from our CMB-PAS dataset and DESI. We show the CMB-PAS results with and without \emph{Planck} low-$\ell$ EE measurements. Without low-$\ell$ EE data, the CMB and BAO measurements are statististically consistent and, hence, can be safely combined. }
    \label{fig:CMB_DESI_OmH0Neff_consistency}
\end{figure*}

\noindent In this appendix, we discuss the impact of the \emph{Planck} low-$\ell$ polarization measurements on our $N_{\rm eff}$ constraints in more detail. Fig.~\ref{fig:CMB_DESI_OmH0Neff_consistency} compares constraints on $H_0r_d$, $\Omega_m$, and $N_{\rm eff}$ from DESI DR2 BAO with the CMB-PAS dataset. In the $H_0r_d-\Omega_m$-plane, the Gaussian tension between DESI DR2 BAO and CMB-PAS is at $3.1\sigma$, making it inappropriate to combine these datasets. However, after removing the low-$\ell$ EE data, the two datasets are consistent within $1.6\sigma.$\footnote{If we evaluate the tension in the full $H_0r_d-\Omega_m-N_{\rm eff}$ space, we find slightly lower values of $2.9\sigma$ and $1.4\sigma$, but we opt to use the more conservative tension metric evaluated in the $H_0r_d-\Omega_m$-plane because this is better representative of the degrees of freedom that DESI BAO measurements actually constrain. We also note that Ref.~\cite{SPT-3G:2025bzu} chose not to combine DESI DR2 BAO with their CMB-SPA dataset because of a $>3\sigma$ discrepancy in the $H_0r_d-\Omega_m-N_{\rm eff}$ plane. We have reproduced this result, noting that the CMB-SPA dataset is in slightly larger tension with DESI than the CMB-PAS dataset considered here because the CMB-SPA combination uses a low-$\ell$ EE likelihood that favors lower values of $\tau$ than \texttt{Sroll2}.} Therefore, in our baseline analysis, we conservatively exclude low-$\ell$ EE data.

\begin{figure*}[!t]
    \centering
    \includegraphics[width=0.95\linewidth]{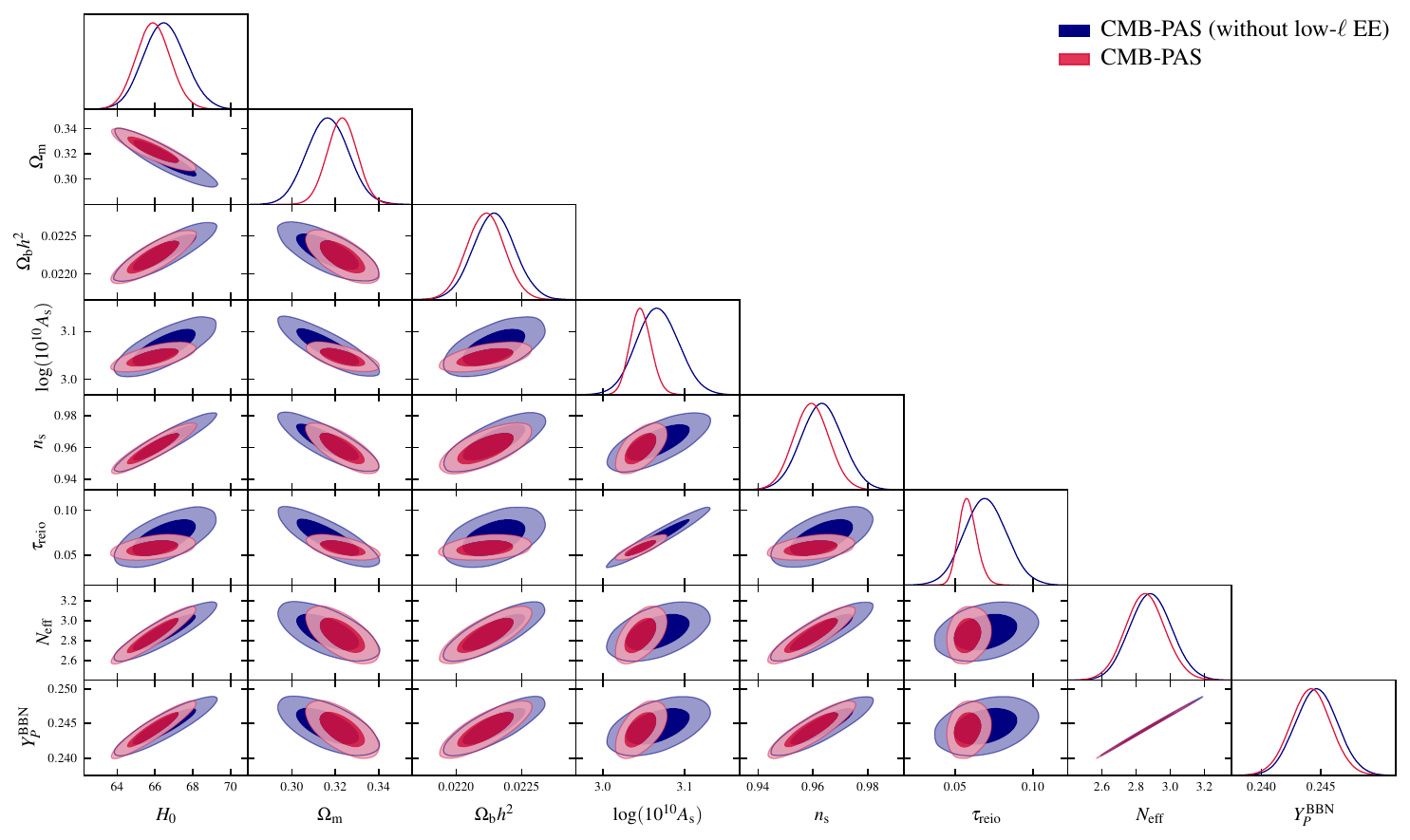}
        \caption{Impact of removing the large-scale \emph{Planck} polarization measurements on the CMB-PAS parameter constraints.}
    \label{fig:full_posterior_nolowlEE_noAbund_noBAO}
\end{figure*}

\begin{figure*}[!b]
    \centering
    \includegraphics[width=0.95\linewidth]{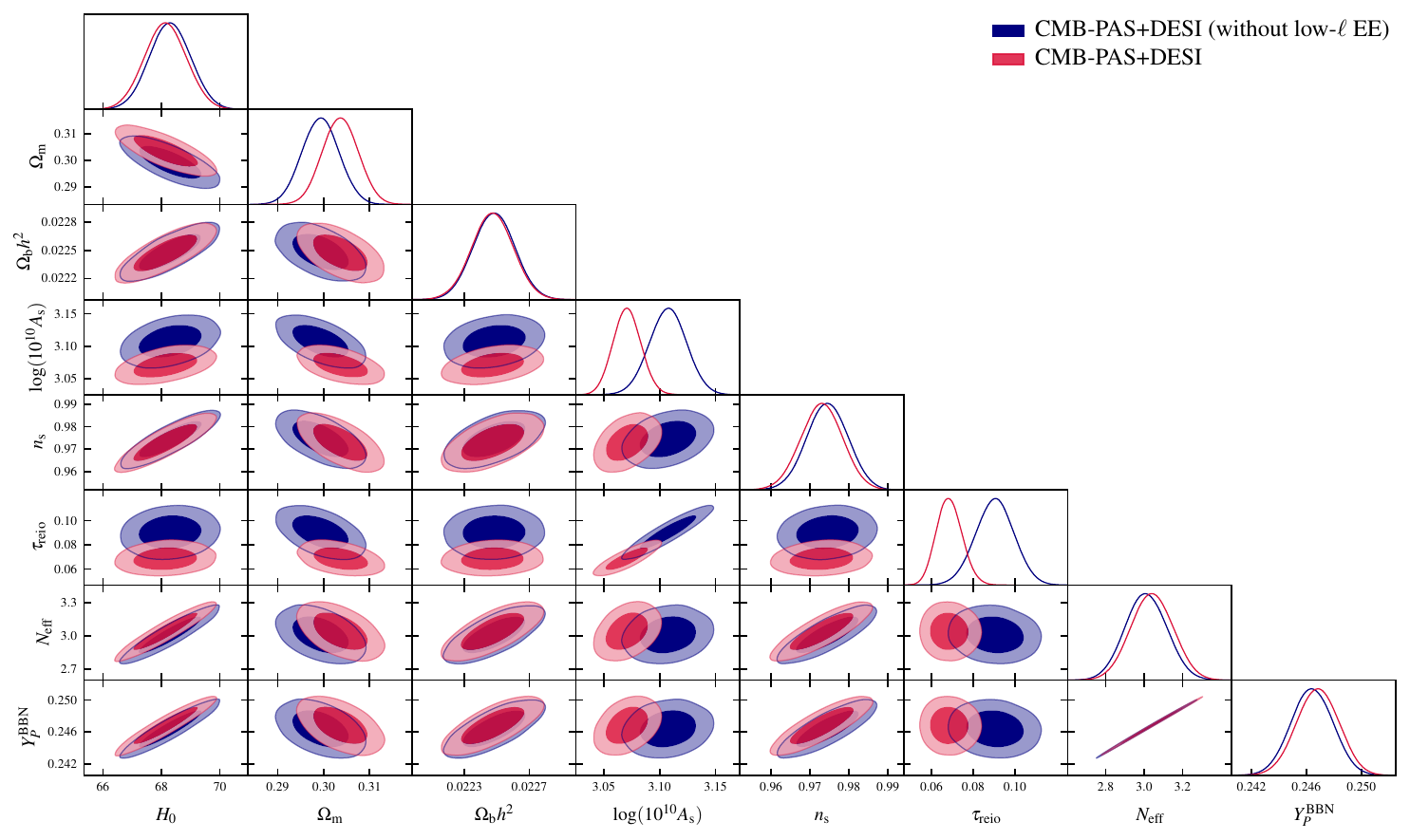}
        \caption{Same as Fig.~\ref{fig:full_posterior_nolowlEE_noAbund_noBAO}, but including DESI DR2 BAO measurements.}
    \label{fig:full_posterior_nolowlEE_noAbund_withBAO}
\end{figure*}

\begin{figure*}[!t]
    \centering
    \includegraphics[width=0.95\linewidth]{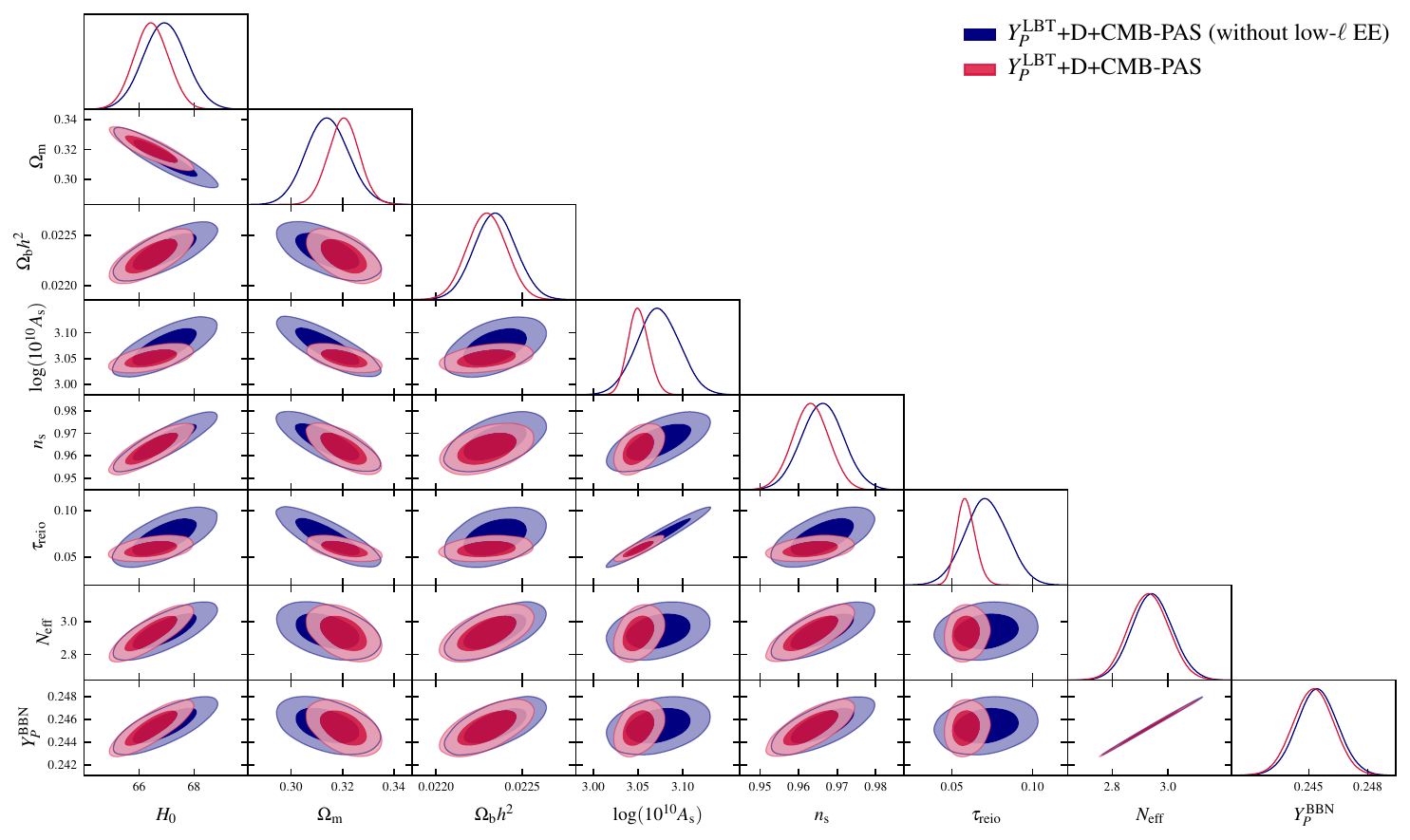}
        \caption{Same as Fig.~\ref{fig:full_posterior_nolowlEE_noAbund_noBAO}, but including primordial abundance measurements. }
    \label{fig:full_posterior_nolowlEE_withAbund_noBAO}
\end{figure*}

\begin{figure*}[!b]
    \centering
    \includegraphics[width=0.95\linewidth]{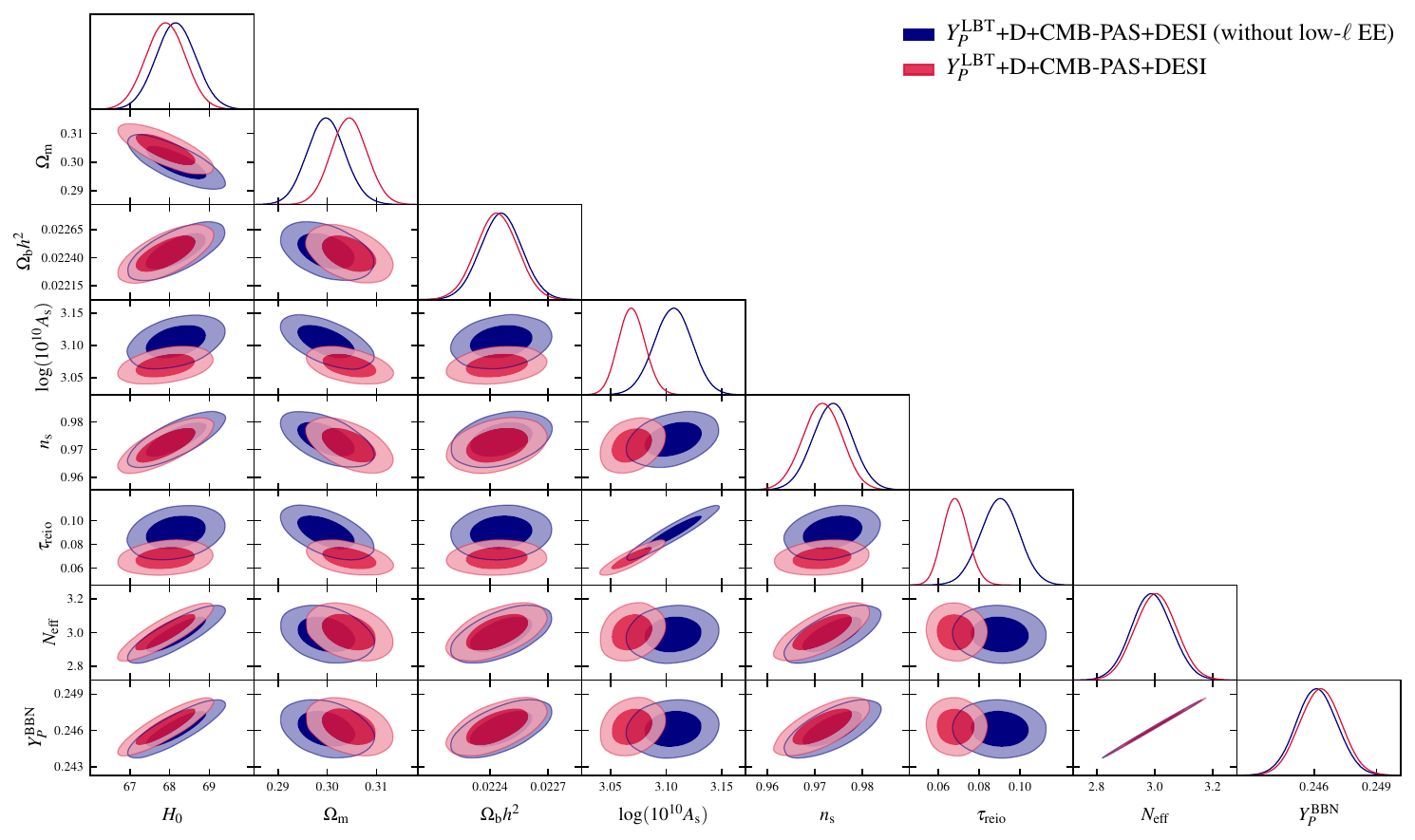}
        \caption{Same as Fig.~\ref{fig:full_posterior_nolowlEE_noAbund_noBAO}, but including primordial abundance and DESI DR2 BAO  measurements.}
    \label{fig:full_posterior_nolowlEE_withAbund_withBAO}
\end{figure*}

In addition to our baseline analysis, we repeat all runs including \emph{Planck} low-$\ell$ EE measurements to assess the overall sensitivity of our $N_{\rm eff}$ constraints to uncertainty in the optical depth to reionization. With the exception of forming a statistically consistent CMB and BAO dataset in an $N_{\rm eff}$ cosmology, there is no reason to remove the low-$\ell$ EE data from our dataset combinations without BAO, \emph{e.g.}, CMB-PAS with \emph{Planck} low-$\ell$ EE constitutes a completely consistent dataset for deriving $N_{\rm eff}$ constraints.
Figs.~\ref{fig:full_posterior_nolowlEE_noAbund_noBAO}-\ref{fig:full_posterior_nolowlEE_withAbund_withBAO} compare the two-dimensional marginalized posterior distributions for the various dataset combinations considered in this work with and without the \texttt{Sroll-2} low-$\ell$ EE likelihood. Whereas the large-scale E-mode measurements are crucial for constraining $\tau$, and, hence, $A_s$ and (to a lesser extent) $n_s$, they have a negligible impact on the $N_{\rm eff}$ constraint.

Table~\ref{table:param_constraints_with_EE} shows the marginalized posterior mean and 68\% two-tailed confidence limit for these dataset combinations including low-$\ell$ EE measurements.

\begin{table}[!h]
\centering
\resizebox{\textwidth}{!}{
\begin{tabular}{|l|c|c|c|c|c|c|c|c|}
\multicolumn{9}{c}{Without abundance information} \\
\hline
Dataset & $H_0$ & $\Omega_m$ & $\Omega_b h^2$ & $\tau$ & $\log A_s$ & $n_s$ & $N_{\rm eff}$ & $Y_P$ \\
\hline
    \emph{Planck} & $66.4 \pm 1.4$ & $0.3183 \pm 0.0092$ & $0.02225 \pm 0.00022$ & $0.0578^{+0.0053}_{-0.0064}$ & $3.044 \pm 0.015$ & $0.9597 \pm 0.0082$ & $2.89 \pm 0.19$ & $0.2446 \pm 0.0027$ \\
    CMB+PAS & $65.90 \pm 0.91$ & $0.3232 \pm 0.0068$ & $0.02222 \pm 0.00014$ & $0.0582^{+0.0051}_{-0.0063}$ & $3.046^{+0.012}_{-0.013}$ & $0.9594 \pm 0.0065$ & $2.85 \pm 0.12$ & $0.2441 \pm 0.0017$ \\
    CMB+PAS+DESI & $68.14 \pm 0.71$ & $0.3036 \pm 0.0039$ & $0.02247 \pm 0.00013$ & $0.0686^{+0.0057}_{-0.0065}$ & $3.071 \pm 0.012$ & $0.9730 \pm 0.0053$ & $3.04 \pm 0.11$ & $0.2468 \pm 0.0015$ \\
\hline
\end{tabular}
}

\vspace{0.5em}

\resizebox{\textwidth}{!}{
\begin{tabular}{|l|c|c|c|c|c|c|c|c|}
\multicolumn{9}{c}{With abundance information} \\
\hline
Dataset & $H_0$ & $\Omega_m$ & $\Omega_b h^2$ & $\tau$ & $\log A_s$ & $n_s$ & $N_{\rm eff}$ & $Y_P$ \\
\hline
    \emph{Planck} & $66.88 \pm 0.74$ & $0.3165 \pm 0.0075$ & $0.02231 \pm 0.00015$ & $0.0581^{+0.0052}_{-0.0063}$ & $3.048^{+0.012}_{-0.013}$ & $0.9625 \pm 0.0050$ & $2.961 \pm 0.082$ & $0.2457 \pm 0.0012$ \\
    CMB+PAS & $66.44 \pm 0.63$ & $0.3205 \pm 0.0060$ & $0.02229 \pm 0.00011$ & $0.0587^{+0.0052}_{-0.0062}$ & $3.051^{+0.011}_{-0.012}$ & $0.9631 \pm 0.0047$ & $2.930 \pm 0.072$ & $0.2452 \pm 0.0010$ \\
    CMB+PAS+DESI & $67.90 \pm 0.49$ & $0.3044 \pm 0.0036$ & $0.02243 \pm 0.00011$ & $0.0684^{+0.0058}_{-0.0064}$ & $3.069 \pm 0.012$ & $0.9715 \pm 0.0041$ & $3.004 \pm 0.071$ & $0.24631 \pm 0.00099$ \\
\hline
\end{tabular}
}
\caption{Same as Table~\ref{table:baseline_param_constraints}, but including \emph{Planck} low-$\ell$ EE data.}
\label{table:param_constraints_with_EE}
\end{table}

\clearpage 


\clearpage


\clearpage

\clearpage

\clearpage\pagebreak

\section{CAMB settings}\label{app:camb_settings}
\noindent In this appendix, we list the \texttt{camb} settings used for all analyses in this work. We adopt the more precise settings from either the ACT DR6 analysis~\cite{ACT:2025tim} or the SPT-3G D1 analysis~\cite{SPT-3G:2025bzu}.
\begin{tcolorbox}[colback=gray!15, colframe=gray!50, title=CAMB settings, sharp corners, boxrule=0.5pt]
\begin{verbatim}
camb:
    extra_args:
      kmax: 10
      k_per_logint: 130
      nonlinear: true
      lens_potential_accuracy: 8
      lens_margin: 2050
      AccuracyBoost: 1
      lAccuracyBoost: 1.2
      lSampleBoost: 1
      min_l_logl_sampling: 6000
      DoLateRadTruncation: false
      lmax: 9000 # Increased from 6000
      recombination_model: CosmoRec
      halofit_version: mead2020
      bbn_predictor: ../PRyMordial/PRyM_Yp_DH_cosmoMC_2023.dat
\end{verbatim}
\end{tcolorbox}

\end{document}